\documentclass[prl,twocolumn,aps,amsmath,amssymb,showpacs]{revtex4}

\usepackage{t1enc}
\usepackage[final]{graphicx}
\usepackage{graphicx}
\usepackage{epsfig}

\begin{document}

\title{Quasi-stationary trajectories of the HMF model: a topological perspective}
\author{Francisco A. Tamarit}
\email{tamarit@famaf.unc.edu.ar}
\altaffiliation{Member of CONICET, Argentina}
\affiliation{ Facultad de Matem\'atica, Astronom\'\i a y F\'\i sica,
         Universidad Nacional de C\'ordoba,
         Ciudad Universitaria, 5000 C\'ordoba, Argentina}
\author{Germ\'an Maglione}
\email{maglione@tero.fis.uncor.edu}
\affiliation{ Facultad de Matem\'atica, Astronom\'\i a y F\'\i sica,
         Universidad Nacional de C\'ordoba,
         Ciudad Universitaria, 5000 C\'ordoba, Argentina}
\author{Daniel A. Stariolo}
\email{stariolo@if.ufrgs.br}
\altaffiliation{Research Associate of the Abdus Salam International Centre
for Theoretical Physics, Trieste, Italy}
\affiliation{Departamento de F\'{\i}sica, 
             Universidade Federal do Rio Grande do Sul,
             CP 15051, 91501-970, Porto Alegre, Brazil} 
\author{Celia Anteneodo}
\email{celia@cbpf.br}
\affiliation{Departamento de F\'{\i}sica, Pontif\'{\i}cia 
Universidade Cat\'olica do Rio de Janeiro, 
CP 38071, 22452-970, Rio de Janeiro, Brazil and \\ 
Centro Brasileiro de Pesquisas F\'{\i}sicas, Rua Dr. Xavier Sigaud 150, CEP 22290-180, 
Rio de Janeiro, Brazil}
\date{\today}


\begin{abstract}

We employ a topological approach to investigate the nature 
of quasi-stationary states   
of the Mean Field XY Hamiltonian model that arise when the system  
is initially prepared in a fully magnetized configuration. 
By means of numerical simulations and analytical considerations, we show that,  
along the quasi-stationary trajectories, the system evolves in a manifold 
of critical points of the potential energy function.  
Although these critical points are maxima,  the large number of directions 
with marginal stability may be responsible for the slow relaxation dynamics  
and the trapping of the system in such trajectories.  

\end{abstract}

\pacs{PACS numbers: 05.20.-y,02.40.-k,64.60.Cn} 

\maketitle


The so called Mean Field $XY$ Hamiltonian Model (HMF model) 
\cite{Antoni} has
received great attention during the last years in the statistical
mechanics community, mainly because of the richness of its dynamical
behavior \cite{Latora01a,anteneodo,all,Montemurro,Pluchino,yamaguchi,superdif}. 
The model is defined by a set of $N$ particles,
or rotators, moving on a unitary circle. The dynamics of the system
is ruled by the following Hamiltonian:

\begin{equation}
{\cal H } = \frac{1}{2} \sum_{i=1}^{N} p_i^2 + \frac{1}{2N} 
\sum_{i,j=1}^N [1- \cos{(\theta_i - \theta_j) }]  \, .
\end{equation} 
Here $\theta_i$ represents the rotation angle of the $i$-th particle
(with $\theta_i \in (0,2\pi])$ and $p_i$ its conjugate momentum. 
This model can be considered as a kinetic
version of the mean field $XY$ magnetic model with ferromagnetic
interactions. From a thermodynamical point of view the model is extremely
simple, in contrast to its rich and still not well understood dynamical behavior. 
On one side, being a mean field model, its equilibrium thermodynamics can be 
exactly solved in the canonical ensemble, yielding a second order 
ferromagnetic phase transition. On the other side its relaxational dynamics
is very complex and equilibrium is not easily attained from an important set
of initial conditions \cite{yamaguchi,superdif,hmf}. 
The origin of this kinetic complexity is elusive and
some similarity with the phenomenology of disordered systems has 
been advocated \cite{Montemurro,Pluchino}.
Nevertheless, the existence of this relation is not at all obvious.
In first place, there is no imposed disorder on the
Hamiltonian. Second, the couplings are all ferromagnetic,
avoiding in this case any kind of structural frustration.
Finally, the infinite range of the interactions further simplifies 
both the dynamics and the thermodynamics of the model, avoiding
any topological consideration on the structure of the lattice
where particles are located. 
On the other hand, and also due to the infinite range of the interactions, 
its dynamics can be efficiently integrated in computer simulations.
Therefore, this is an excellent prototype for analyzing the
microscopic dynamics of a finite system close to a critical 
point.

The microcanonical simulations of the HMF reproduce many of the
anomalous critical behaviors observed in nuclear and cluster fragmentation
processes. 
In particular, depending on the initial preparation of the system, it is possible
to verify the existence of negative specific  heat curves, qualitatively
similar to those observed in recent fragmentation experiments
in small clusters  (see for instance \cite{experiment} and references therein).
But the interest on this model largely exceeds this
motivation. 
The existence of quasi-stationary solutions whose
life-times diverge \cite{Latora01a,yamaguchi}, in the thermodynamical 
limit $N\to \infty$, 
has raised the question on whether it is possible or not to construct
a measure theory able to predict the stationary values
of physical observables in long standing out of equilibrium regimes
\cite{Tsallis88}.
Furthermore, the existence of a glassy-like relaxation dynamics 
\cite{Montemurro,Pluchino} along
quasi--stationary trajectories, has opened new challenging
questions on the origin of such unexpected behavior for  an
unfrustrasted non disordered model. 
Finally, in the
last few years, the HMF model has been used, as a paradigmatic
example, for the study of the so called {\em Topological Hypothesis}
\cite{geometric,Casetti03,theorem} which asserts 
that phase transitions, {\em even in a
finite system}, can be identified by searching for drastic
topological changes in the submanifolds of the 
interaction potential.

Concerning the HMF thermodynamics \cite{Antoni}, the usual microcanonical 
and canonical calculations
predict that the system suffers a second order phase transition
at $T_c = 1/2$ (which corresponds to an internal energy per
particle $U_c/N = 3/4$), 
from a low temperature ordered phase to a high temperature
disordered one. Actually, one can associate to each
rotator a local magnetization:
$\vec{m}_i = (\cos{\theta_i}, \sin{\theta_i} ),$
and then define the order parameter of the transition as
the global magnetization:  
\begin{equation}
\vec{M} = (M_x, M_y)=  \frac{1}{N} \sum_i \vec{m}_i  .
\end{equation}
At the critical energy, $M\equiv|\vec{M}|$ vanishes continuously  as the
system is heated from the ordered phase.

Concerning the dynamics, when the system is 
prepared very far from equilibrium, 
for energies just below the critical one, the system
gets trapped into quasi-equilibrium trajectories.  
These trajectories are characterized by time averages of 
one-time observables
which reach, after a rapid initial transient, almost
constant values which do not coincide with those  
predicted by microcanonical or canonical ensemble 
calculations \cite{Latora01a}. 
The average time that a system of size $N$ 
remains in a quasi--stationary
trajectory grows with $N$ \cite{Latora01a,yamaguchi}. 
Therefore,  if the system
were infinite, it would remain there forever, without
ever reaching true equilibrium. An even more surprising
scenario appears when one considers the relaxation
of the two--time correlation function $C(t,t^{\prime})$
(either in the whole phase space \cite{Montemurro} or
considering only the momenta space \cite{Pluchino}). The
explicit dependence of $C$ on both times $t$ and $t^{\prime}$
indicates the loss of time--translational invariance, 
proper of equilibrium states, and the appearance of  
memory effects, a phenomenon
usually called {\em aging}. The scaling law of the two--time
autocorrelation functions \cite{Montemurro,Pluchino} 
is qualitatively similar to that observed 
in some real spin glasses \cite{Vincent02}.  Nevertheless we will
show that the physical mechanisms behind the quasi-stationary states of
the HMF are completely different from those present in disordered systems.

In this work we will show,
through  numerical simulations, 
that the complex nonequilibrium quasi--stationary regime observed
just below the critical energy can be interpreted from a topological 
point of view.  
Our analysis will focus on the topology of the surface
defined by the potential energy in the configuration
space. The HMF potential energy, as well as any Curie-Weiss like potential, 
can be written in terms of the order parameter of the system: 
\begin{equation}  
V    = \frac{N}{2}(1-M^2) \;.
\label{potentialmf}
\end{equation}
Note that the potential energy per particle $V/N$ 
takes values in the interval 
$0 \le V/N \le v_c\equiv 1/2 $. 
The lower limit corresponds to the case of the fully ordered
configurations (hence $M=1$) and the upper bound to  a
completely disordered configuration. 
The configuration space manifold $\cal M$
is an $N$-dimensional torus parametrized by the $N$ angles $\theta_i$. 
The critical points (CPs) of $\cal M$ are those points for which all
the $N$ derivatives of $V/N$ vanish, i.e., 
${\partial (V/N)}/{\partial \theta_i } = 0$, for $i=1,\ldots,N$. 
Making use of the infinite range of the interactions, one can
write the derivatives of the potential in terms of the two components
of the order parameter, namely, 
\begin{equation}
\frac{\partial V/N}{\partial \theta_i } = 
\frac{1}{N}M_x\sin{\theta_i} - \frac{1}{N}M_y \cos{\theta_i} = 0 \;.
\label{derivatives}
\end{equation}

Furthermore, CPs can be classified according to 
the  eigenvalues of the Hessian of $V/N$, 
that for the HMF can be written as $\bf  H = D+ B$ \cite{Casetti03}, 
where 

\begin{eqnarray} \nonumber
B_{ij} &=& -\frac{1}{N^2}\ (1-\delta_{ij})\ \cos({\theta_i- \theta_j})  
- \frac{1}{N^2}\ \delta_{ij} \,, \\
\label{hessian}
D_{ij} &=& \frac{\delta_{ij}}{N}\ (M_x\cos{\theta_i} + M_y \sin{\theta_i}) \, .
\end{eqnarray}

In our work we
use the following protocol: starting from
a far-from-equilibrium configuration, we integrate
numerically the set of Hamilton equations of the system using
a fourth order symplectic method with a very small time step
(typically $dt=0.01$). Along the trajectories
we evaluate, at each time step, the modulus of the $N$  
derivatives of $V/N$ given by (\ref{derivatives}) and identify
the maximum over $i=1,\ldots,N$, through 
\begin{equation}
\lambda = N\max_{i} \left| \frac{\partial (V/N)}{\partial \theta_i}
\right| \, .
\end{equation}
Then, each time that $\lambda = 0$, it means that the system reaches a CP 
(actually, due to the finiteness of time step, $\lambda$ is
never exactly zero, but it gets closer as $dt$ decreases).

\begin{figure}[ht]
\begin{center}
\includegraphics*[bb=50 250 590 705, width=0.45\textwidth]{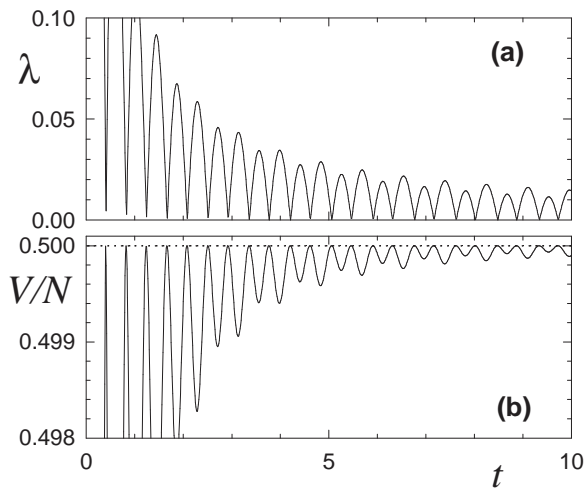} 
\caption{ Time evolution of (a) $\lambda$, the largest modulus of the derivatives of $V$, 
and (b) $V/N$, the potential energy per particle, 
for a system of $500$ particles and $U/N=10$, well above the
phase transition.  The system was initially prepared in a 
regular water-bag configuration. The dotted line in (b) corresponds to 
the equilibrium value in the thermodynamic limit.}
\label{fig:lambda10}
\end{center}
\end{figure}

Let us first analyze the behavior of the system in the disordered
phase. It is important to stress that,
although the system ultimately relaxes to equilibrium, 
a slow relaxation has been observed
when starting very far from equilibrium \cite{superdif}. 
This is probably due to the 
fact that rotators move almost freely and trajectories are weakly  chaotic \cite{anteneodo,lyaps}.  
In Fig. \ref{fig:lambda10}, we plot $V/N$ and $\lambda$
as a function of $t$, for $U/N=10$, in the disordered phase well above the 
critical energy $U_c/N=3/4$.  The system has been initially prepared in a ``water-bag'' 
configuration, with all the rotators aligned along the $x$ axis
($\theta_i = 0$, for all $i$) and the momenta drawn from a
uniform distribution (actually we used regularly spaced momenta \cite{vlasov}). 
Fig. \ref{fig:lambda10} indicates that the system periodically visits 
CPs of the potential, corresponding to $V/N=1/2$ (hence $M=0$).  
In fact the observed period is of the order of the mean period of rotation \cite{lyaps}.   

\begin{figure}[ht]
\begin{center}
\includegraphics*[bb=50 250 590 705, width=0.45\textwidth]{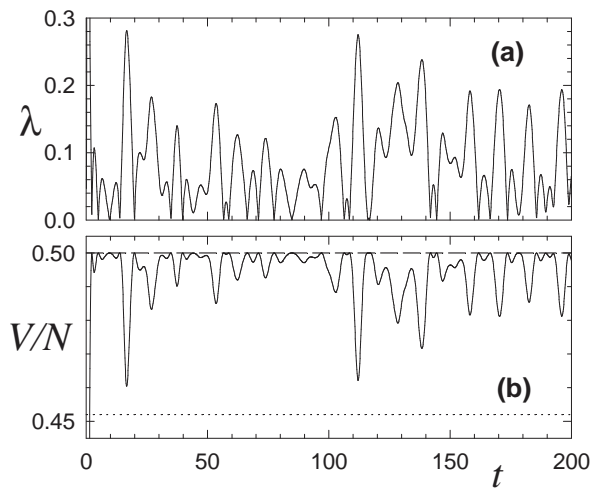}
\caption{Time evolution of $\lambda$ (a) and $V/N$ (b)
for a system of $500$ particles and $U/N=0.69$, just below the
phase transition. The system was initially prepared in a 
regular water-bag configuration. 
The dotted and dashed lines in (b) correspond to 
equilibrium and metaequilibrium values in the thermodynamic limit, respectively.}
\label{fig:lambda069}
\end{center}
\end{figure}

\begin{figure}[ht]
\begin{center}
\includegraphics*[bb=90 300 495 570, width=0.45\textwidth]{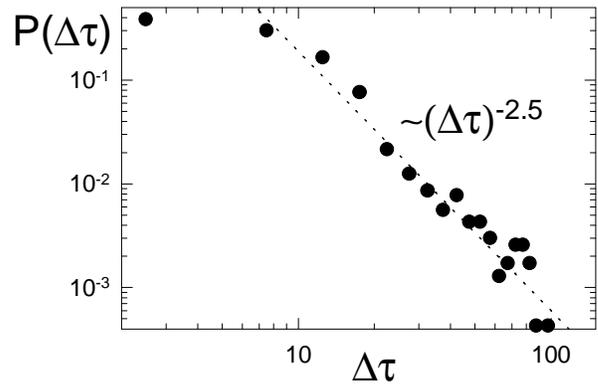} 
\caption{Probability distribution function $P(\Delta \tau)$ of the time intervals
between two consecutive critical points, for the
same system of Fig. \ref{fig:lambda069}.}
\label{fig:pdftau}
\end{center}
\end{figure}

Let us now compare these results with those obtained in the
low energy phase, just below the phase transition. 
In this case, 
quasi-stationary solutions emerge, displaying a kind of
glassy-like dynamics characterized by weak chaos, non-Gaussian 
velocity distributions and sub--aging, as mentioned above.
In Fig.~\ref{fig:lambda069}, we plot $V/N$ and $\lambda$  vs. $t$ 
for $U/N=0.69$ and water-bag initial conditions. 
Here we verify that again the system sequentially visits
one CP of the potential energy after another, also corresponding to 
$M=0$. 
However, at variance with the high energy phase, the time intervals  
$\Delta \tau$ elapsed between two
successive CPs do not present any pattern of
periodicity. On the contrary, the system visits the CPs in an apparently
disordered way. 
The probability distribution function (PDF)
$P(\Delta \tau)$ of time intervals between CPs is shown in Fig.~\ref{fig:pdftau}.
$P(\Delta \tau)$ can be reasonably fitted by a power law decay.
Other striking features of the dynamical behavior can be noted
in Fig.~\ref{fig:lambda069}. First, the system is initialized in a configuration
at the bottom of the potential energy $V/N=0$ and it goes almost abruptly to
a region with a mean potential energy per particle larger than the equilibrium
mean potential energy, and stays around this level during the whole time span
of the simulation. In fact it goes close to the top of the potential energy landscape,
$V/N\approx 1/2$, tries to escape downhill but uses the kinetic energy gain to 
attain again CPs at the top. Somehow the system is not able to relax
from this level to the equilibrium level during very large time scales. The
configurations sampled correspond to the quasi-stationary states and the system
stays there during time spans which scale with the size N, being trapped forever
in the thermodynamic limit. One can say that, in configuration space, 
quasi-stationary states are always near CPs of the landscape.

We next address the effect of system size. 
The same qualitative behavior can be observed,
with a bigger numerical effort, for larger systems. 
The departure from the CPs (measured by the height of
$\lambda$ between consecutive CPs) decreases as
$N$ increases, but the distribution of $\Delta\tau$ 
remains unaltered.   
From Eqs. (\ref{potentialmf}) and (\ref{derivatives}), it is straightforward
to see that at the critical level $v_c=1/2$ the CPs are 
continuously degenerate. Consequently as N grows and fluctuations outside
the critical level diminish the system wanders more and more inside this
manifold of CPs. 
As the energy decreases, the CPs are visited more sparsely,  
down to $U/N\approx 0.67$, where the system stops wandering among CPs.  
It is noteworthy that spatially homogeneous states lose 
Vlasov stability approximately at that energy \cite{vlasov}. 

\begin{figure}[ht]
\begin{center}
\includegraphics*[bb=50 250 590 600, width=0.45\textwidth]{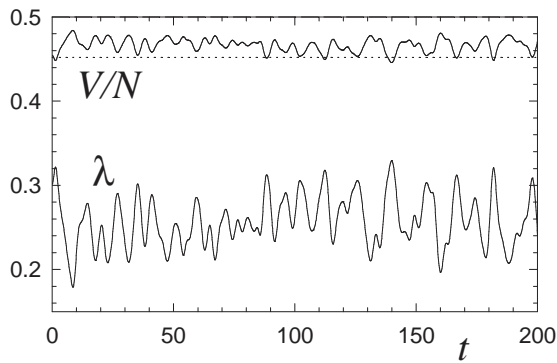}
\caption{Time evolution of $\lambda$ and $V/N$  
for a system of $500$ particles and $U/N=0.69$, just below the
phase transition. The system was initially prepared in a close to equilibrium 
configuration. The dotted line corresponds to 
the equilibrium value of $V/N$.}
\label{fig:lambdaeq}
\end{center}
\end{figure}

A completely different scenario
emerges when the system is prepared, at a given energy, in
an almost equilibrated initial configuration. The plot
of $\lambda$ vs. $t$  presented in Fig. \ref{fig:lambdaeq}  clearly indicates
that, in this case, the system {\em does not wander} among CPs. 
Its potential energy per particle (and then also
its temperature) rapidly starts fluctuating around their
canonical values, showing strong finite size effects.

A crucial piece of information comes from 
the stability of the visited CPs. 
Since capturing the exact time at which the 
dynamics passes through a CP is a difficult task, numerical evaluation of the 
Hessian at CPs arising from the dynamics may lead to wrong estimates. 
Therefore it is important to perform some analytical calculations. 
In order to do so, let us recall that, both in 
the high energy phase and in the quasi-stationary states, 
the CPs correspond to $V/N=1/2$ (see Figs. 1 and 2), hence they are points of 
$\cal M$ with zero magnetization.  
Moreover, the distribution of angles at those points is approximately uniform \cite{vlasov}. 
A configuration with these characteristics for which analytical calculations are possible, 
consists in regularly distributed angles in the interval $(0,2\pi]$, i.e., 
$\theta_k=2\pi k/N$, $k=1,...,N$ for even $N$. 
If $M=0$, from Eq. (\ref{hessian}), the Hessian of the potential energy is $\bf H=B$. 
Then, for the regular configuration  we have

\begin{equation}
{\bf H} = {\bf B} =-\frac{1}{N^2}\ (\openone +{\bf A} ),
\end{equation}
where $A_{kl}=\cos(2\pi(k-l)/N)$. 
This circulant matrix can be diagonalized in Fourier space, yielding the following 
eigenvalues of the Hessian matrix
\begin{equation}
H_l = -\frac{1}{N^2}\biggl(  \cos(\pi l) +2\sum_{r=0}^{N/2}\cos(2\pi rl/N)\cos(2\pi r/N)  
\biggr),
\end{equation}
for $1\leq l\leq N$. 
Thus, we obtain $H_1=H_{N-1}=-1/(2N)$ while the remaining eigenvalues are null. 
This means that at these CPs there are two unstable directions and $N-2$ marginal ones. 
This picture remains valid for more general situations than those restricted to the particular 
regular cases. In fact, we have verified that when angles are randomly chosen in 
the interval $(0,2\pi]$ there are two eigenvalues with values close to $-1/(2N)$ and the 
remaining $N-2$ vanish. 
Finally, the eigenvalues calculated from the configurations of the dynamics very 
close to CPs are also consistent with this picture. 
>From this analysis it is clear that
in the high energy phase and also in the quasi-stationary low energy states, the systems
wanders in an almost flat landscape. 

In the high energy phase, energy is mainly kinetic, leading to ballistic 
behavior in the flat landscape of CPs during long time scales. 
In the low energy, quasi-equilibrium regime, kinetic energy is comparable to the 
potential one. At this relatively low energies diffusion is slower than ballistic.
Nevertheless, due to the flatness of configuration space 
superdiffusive behavior is observed \cite{superdif}. 

Summarizing, we have seen, by means of numerical
simulations, that the quasi-stationary trajectories
observed in the HMF model can be interpreted in terms
of the topological properties of the potential
energy per particle $V/N$ of the model. Starting from a
far from equilibrium configuration with $V/N=0$, the
system initially decreases as much as possible its
kinetic energy and settles at the {\em flat} top of its potential
energy. 
The simulations confirm that along
the  quasi stationary states the system wanders among different
CPs with only two negative directions and $N-2$ 
marginal ones.  Moreover, CPs
at the upper critical level are continuously degenerate. This suggests that
once inside this critical submanifold the system cannot go out easily and
relax to equilibrium. 
A probable scenario in the thermodynamic limit 
is that, as the fluctuations of the potential energy close to the 
critical level $v_c=1/2$ go to zero,
the system keeps wandering continuously inside the manifold of CPs 
and consequently remains forever out of thermodynamic equilibrium.

This work was partially supported by 
CONICET (Argentina), Agencia Córdoba Ciencia (Argentina),
Secretar\'{\i}a de Ciencia y Tecnolog\'{\i}a de la Univ. Nac.
C\'ordoba (Argentina) and CNPq (Brazil).


%

\end{document}